\documentclass{PoS}
\usepackage{cite}
\usepackage{amsmath}
\usepackage{amssymb}
\usepackage{epsfig}
\usepackage{etex, xy}           
\xyoption{all}

\newcommand{\gsim}{\raisebox{-0.07cm   }
{$\, \stackrel{>}{{\scriptstyle\sim}}\, $}}

\title{{\footnotesize \rm DESY 16/002, DO-TH 16/02, MITP/16-016 
}\\
3-Loop Corrections to the Heavy Flavor Wilson Coefficients in Deep-Inelastic Scattering\thanks{This work was 
supported in part by the Austrian Science Fund (FWF) grants P 27229, SFB F50 (F5009-N15), FP7 
ERC Starting Grant 257638 PAGAP and the European Commission through contract PITN-GA-2012-316704 ({HIGGSTOOLS}).}}

\ShortTitle{3-Loop Corrections to the Heavy Flavor Wilson Coefficients}

\author{
J.~Ablinger$^1$, A.~Behring$^2$, \speaker{J.~Bl\"umlein}$^2$,
A.~De~Freitas$^2$, A. Hasselhuhn$^3$, A.~von~Manteuffel$^4$, C.G.~Raab$^{1,5}$, 
M.~Round$^{1,2}$, C.~Schneider$^1$, and F.~Wi\ss{}brock$^{1,2,6}$ 

{\it \small $^1$~Research Institute for Symbolic Computation (RISC),
                          Johannes Kepler University,\\ ~~~Altenbergerstra\ss{}e 69,
                          A--4040, Linz, Austria \\
$^2$~Deutsches Elektronen-Synchrotron, DESY, Platanenallee 6, D-15738 Zeuthen, Germany}\\
        ~~~E-mail: \email{johannes.bluemlein@desy.de} \\
{$^3$ Institut f\"ur Theoretische Teilchenphysik,
Karlsruher Institut f\"ur Technologie (KIT),\\
~~~D--76128 Karlsruhe, Germany} \\
{$^4$ PRISMA Cluster of Excellence,  Institute of Physics, J. Gutenberg
University, D--55099 Mainz,\\  ~~~Germany.}
\\
{$^5$ Johann Radon Institut (RICAM), 
Altenbergerstra\ss{}e 69, A-4040 Linz, Austria}\\
{$^6$ IHES, 35 Route de Chartres, F--91440 Bures-sur-Yvette, France}}

\abstract{A survey is given on the status of 3-loop heavy flavor corrections to deep-inelastic 
structure functions at large enough virtualities $Q^2$.}

\FullConference{The European Physical Society Conference on High Energy Physics\\
                 22-29 July 2015\\
                 Vienna, Austria}

\begin{document}

\section{Introduction}

\vspace*{1mm}
\noindent
The determination of fundamental parameters of the Standard Model such as the strong coupling constant $\alpha_s(M_Z^2)$ 
\cite{ALPH} and the mass of the charm quark $m_c$ \cite{Alekhin:2012vu} from the precision World data on unpolarized 
deep-inelastic lepton-nucleon scattering requires the complete 3-loop QCD corrections. While the massless 3-loop 
corrections are known \cite{Moch:2004pa,Vogt:2004mw,Vermaseren:2005qc}, the calculation of the massive corrections at higher 
values of 
$Q^2/m_c^2 \gsim 10$ is underway \cite{Blumlein:2014zxa}. Here $q$ denotes the 4-momentum transfer to the hadronic system and 
$Q^2 = -q^2$. 
The 2-loop corrections are known in semi-numerical form
\cite{Laenen:1992zk}\footnote{For an implementation in Mellin space, see \cite{Alekhin:2003ev}.} and for large 
values of $Q^2$, $Q^2/m_c^2 \gsim 10$ also in analytic form \cite{Buza:1995ie,Bierenbaum:2007qe}. A series of Mellin 
moments of the 3-loop massive corrections has been computed in Ref.~\cite{Bierenbaum:2009mv} for the operator 
matrix elements (OMEs) and Wilson coefficients contributing to the structure function $F_2(x,Q^2)$
mapping the problem to massive tadpole-structures, which have been calculated using {\tt MATAD} 
\cite{Steinhauser:2000ry}. All logarithmic 
terms to 3-loop order have been calculated both for the massive Wilson coefficients as well as for the 3-loop 
transition matrix elements in the variable flavor number scheme (VFNS) in Ref.~\cite{Behring:2014eya}, referring
also to the massive 2-loop OMEs up to $O(\varepsilon)$ \cite{Bierenbaum:2008yu,Bierenbaum:2009zt}, the 3-loop 
anomalous dimensions \cite{Moch:2004pa,Vogt:2004mw}, and the 2-loop massless Wilson coefficients \cite{ZN}. At 3-loop order all
$T_F^2 N_F$-terms have been calculated \cite{Ablinger:2010ty}, as well as the $T_F^2$-terms for the gluonic
OMEs \cite{Ablinger:2014uka}. The asymptotic 3-loop contributions to the structure function $F_L(x,Q^2)$ were 
calculated in \cite{Blumlein:2006mh,Behring:2014eya} and charged current corrections up to 2-loop order
were given in \cite{CC}. In a series of technical papers we presented details of the calculation technologies used 
\cite{Ablinger:2012qm,Ablinger:2014yaa,Ablinger:2015tua} and the properties of mathematical structures occurring
\cite{Blumlein:1998if,Vermaseren:1998uu,Blumlein:2003gb,Blumlein:2009ta,Blumlein:2009fz,Blumlein:2009cf,
Blumlein:2009tj,Blumlein:2010zv,Moch:2001zr,Ablinger:2013cf,Ablinger:2014bra,Ablinger:2011te,SURV}.

In this note a survey is given on the status of the 3-loop heavy flavor corrections to deep-inelastic structure 
functions, with emphasis on recent developments. There are five different contributions to the structure functions 
$F_{2,L}(x,Q^2)$ from 3-loop order onward. So far four of them have been completed and all the iterative-integral
(over general alphabets) contributions to the Wilson coefficient $H_{2,g}^{(3)}$ in the contributing master integrals 
have been calculated. In the VFNS, eight different OMEs are relevant (counting also transversity), out of which seven 
have been completed. Here most recently we obtained a representation of the Mellin moments for general even integer 
values of $N$ of the OME $A_{gg}^{(3)}$ \cite{AGG}.

After a brief introduction into the formalism, we discuss a series of aspects of the calculation of master 
integrals and present then numerical results on the different heavy flavor Wilson coefficients and OMEs 
contributing to the structure function $F_2(x,Q^2)$, \cite{Ablinger:2014vwa,Behring:2015roa,Behring:2015zaa,
Ablinger:2014lka,Ablinger:2014nga}. 
While the former results presented refer to the case of $N_F$ 
massless and a single heavy quark, we also briefly comment on some analytic results for two massive flavors. 
\section{Basic Formalism}

\vspace*{1mm}
\noindent
In the leading-twist approximation the deep-inelastic structure functions $F_{2,L}(x,Q^2)$ factorize into 
the non-perturbative parton distribution functions $f_j$ and the Wilson coefficients 
$\mathbb{C}_{j,(2,L)}$, which can be calculated perturbatively \cite{Blumlein:2012bf},
\begin{eqnarray}
F_{(2,L)}(x,Q^2)=\sum_j 
\quad {\underbrace{\mathbb{C}_{j,(2,L)}\left(x,\frac{Q^2}{\mu^2},\frac{m^2}{\mu^2}\right)}_{perturbative}} 
\quad\otimes\quad {\underbrace{f_j(x,\mu^2)}_{nonpert.}}.
\end{eqnarray}
Here $x$ denotes the Bjorken variable $x = Q^2/S y$, with $S$ the cms energy squared and $y = 2P.q/S$ the inelasticity, $P$ the 
proton 4-momentum, $\mu$ denoting both the renormalization and factorization scale, and  $\otimes$ denotes the Mellin convolution
\begin{eqnarray}
f(x)\otimes g(x)\equiv \int_0^1 dy\int_0^1 dz\; \delta(x-yz)f(y)g(z).
\end{eqnarray}
The Wilson coefficients contain massless $(C_{j,(2,L)})$ and massive ($H_{j,(2,L)}$) contributions, written 
here 
in Mellin space
\begin{eqnarray}
\mathbb{C}_{j,(2,L)}\left(N,\frac{Q^2}{\mu^2},\frac{m^2}{\mu^2}\right)
= 
{C_{j,(2,L)}\left(N,\frac{Q^2}{\mu^2}\right)}+H_{j,(2,L)}\left(N,\frac{Q^2}{\mu^2},\frac{m^2}{\mu^2}\right).
\end{eqnarray}
At large scales $Q^2$ the massive contributions
\begin{eqnarray}
H_{j,(2,L)}\left(N,\frac{Q^2}{\mu^2},\frac{m^2}{\mu^2}\right)=\sum_{i} 
{C_{i,(2,L)}\left(N,\frac{Q^2}{\mu^2}\right)}{A_{ij}\left(\frac{m^2}{\mu^2},N\right)}
\end{eqnarray}
factorize \cite{Buza:1995ie} into the massless Wilson coefficients and the massive OMEs $A_{ij}$. We are 
performing the analytic calculation of these quantities and obtain the massive Wilson coefficients at large 
enough scales $Q^2$ to $F_2(x,Q^2)$, for $Q^2/m^2 \gsim 10$ at the 1\% level. The explicit representation of the   
Wilson coefficients and the transition formulae in the VFNS for $N_F \rightarrow N_F + 1$ massless quarks to 
3-loop order has been given in Ref.~\cite{Bierenbaum:2009mv}. The OMEs having been calculated so far allow to 
represent 
\begin{eqnarray}
{f_k(N_F+1, \mu^2) + f_{\overline{k}}(N_F+1, \mu^2)}
&=& {A_{qq,Q}^{\rm NS} \Big(N_F, \frac{\mu^2}{m^2}\Big)}\otimes
  {\left[f_k(N_F, \mu^2) + f_{\overline{k}}(N_F, \mu^2)\right]} \nonumber\\
& & + {\tilde A_{qq,Q}^{\rm PS}\Big(N_F, \frac{\mu^2}{m^2}\Big)}\otimes
  {\Sigma(N_F, \mu^2)}
 + {\tilde A_{qg,Q}^{\rm S}\Big(N_F, \frac{\mu^2}{m^2}\Big)}\otimes
  {G(N_F, \mu^2)}
\nonumber\\
\\
{G(N_F+1, \mu^2)} &=&  {A_{gq,Q}^{\rm S}\Bigl(N_F,\frac{\mu^2}{m^2}\Bigr)}
\otimes {\Sigma(N_F,\mu^2)}   
+ {A_{gg,Q}^S\Bigl(N_F,\frac{\mu^2}{m^2}\Bigr)}\otimes {G(N_F,\mu^2)}\nonumber\\
\end{eqnarray}
the transition in the non-singlet case and for the gluon distribution to 3-loop order. Here, $f_k$ denotes the $k$th 
massless quark densities, $\Sigma = \sum_k (f_k + \bar{f}_k)$ the singlet- and $G$ the gluon distribution function.

We turn now to the analytic calculation of the OMEs. 
\section{Calculation of the Master Integrals}
\label{sec:3}

\vspace*{1mm}
\noindent
Due to the large number of integrals being associated to the Feynman integrals contributing to the eight 
principal OMEs, we use the integration-by-parts relations (IBPs) \cite{IBP} implemented in the package {\tt Reduze2}
\cite{Reduze1,Reduze2} and reduce them to master integrals. An overview on the respective numbers is given in the 
following Table.
\begin{center}
\begin{tabular}{|l|r|r|r|}
\hline
& Diagrams & Integrals & Masters \\
\hline
$A_{qq,Q}^{{\rm NS[TR]} (3)}$ &  110 &  5636 &  35 \\
$A_{gq,Q}^{(3)}$          &   86 & 12253 &  41 \\
$A_{Qq}^{{\rm PS} (3)}$   &  125 &  4824 &  66 \\
$A_{gg,Q}^{(3)}$          &  642 & 68131 & 139 \\
$A_{Qg}^{(3)}$            & 1233 & 34123 & 340 \\
\hline
\end{tabular}
\end{center}

For the calculation of the master integrals different analytic technologies are used, see 
Refs.~\cite{Ablinger:2015tua,Ablinger:2016pbw}. The simplest topologies can be represented by
generalized hypergeometric functions and their extensions \cite{HYP} and expanded in the dimensional 
parameter $\varepsilon$ leading to nested finite and infinite sums. In more involved cases, Mellin-Barnes 
representations \cite{MB} need to be used in addition, which also lead to multiple sum representations. 
In case of Feynman diagrams without singularities in $\varepsilon$ also the method of hyperlogarithms 
\cite{Brown:2008um,FWTHESIS,Panzer:2014gra} can be directly applied. The next general class of cases can be solved using the 
method of differential equations \cite{DEQ} for linear systems of master integrals. In the present calculation 
one auxiliary  parameter emerges due to the formal resummation of the local operator insertion into propagator 
terms, the simplest example being
\begin{eqnarray}
\sum_{n=0}^\infty x^n (\Delta.p)^n \rightarrow  \frac{1}{1 - x \Delta.p}~.
\end{eqnarray}
In Refs.~\cite{Ablinger:2015tua,Ablinger:2016pbw} we have developed an algorithm for solving these systems 
in the form they appear in the calculation with no need to choose a special basis \cite{Henn:2013pwa}. 
The differential system is cast into a higher order difference equation by using a decoupling algorithm \cite{DECOUP}
and then solved using the methods 
encoded in the package {\tt Sigma} \cite{SIG1,SIG2}. The remaining equations of the system are solved 
subsequently as linear equations. The corresponding initial values are obtained for fixed Mellin 
moments and are obtained from the equations given by the IBPs. In some cases it is advantageous to 
compute the given multiple integrals using the (multivariate) Almkvist-Zeilberger Theorem \cite{AZ} 
as integration method.

Most of the above methods lead to nested finite and infinite sums after expanding in the dimensional parameter
$\varepsilon$, which again have to be solved using modern summation technologies 
\cite{Karr:81,Schneider:01,Schneider:05a,Schneider:07d,Schneider:10b,Schneider:10c,Schneider:15a,Schneider:08c,DFTheory}, 
encoded in the packages {\tt Sigma} \cite{SIG1,SIG2}, {\tt EvaluateMultiSums, SumProduction} \cite{EMSSP},
{\tt SolveCoupledSystem} \cite{Ablinger:2015tua}, {\tt $\rho$SUM} \cite{RHOSUM}, and {\tt MultiIntegrate} 
\cite{Ablinger:PhDThesis}. Moreover, mutual use is made of the package 
{\tt HarmonicSums} \cite{Ablinger:PhDThesis,Harmonicsums,Ablinger:2011te,Ablinger:2013cf,Ablinger:2014bra}.
The differential equation method leads to sequential difference equations in a different way, which are also solved
using difference-field theory.

Using these methods seven out of eight OMEs could be completely calculated. In fact we could calculate all master integrals
also contributing to $A_{Qg}^{(3)}$, that lead to iterative integrals over whatever alphabet. There remain 105 
master integrals
containing non-iterative integral parts, forming classes of a new kind.
\section{Numerical Results}

\vspace*{1mm}
\noindent
In the following we would like to present some numerical results for the massive Wilson coefficients contributing to
the structure function $F_2(x,Q^2)$ at larger values of $Q^2$ up to 3-loop order in the strong coupling 
constant $a_s = \alpha_s/(4\pi)$ and for  OMEs in the VFNS.

In Figure~1 we we illustrate the contributions up to $O(a_s^2)$ and $O(a_s^3)$ to the structure function 
$F_2(x,Q^2)$ referring to the parton distribution functions \cite{Alekhin:2013nda} here and in the following 
and $m_c = 1.59$~GeV in the on-shell scheme \cite{Alekhin:2012vu}.
It turns out that the 2-loop corrections are even smaller than those at 3-loops because of a small $x$ effect 
contributing from 3-loops onward.
\begin{figure}[t]
\includegraphics[width=0.48\textwidth]{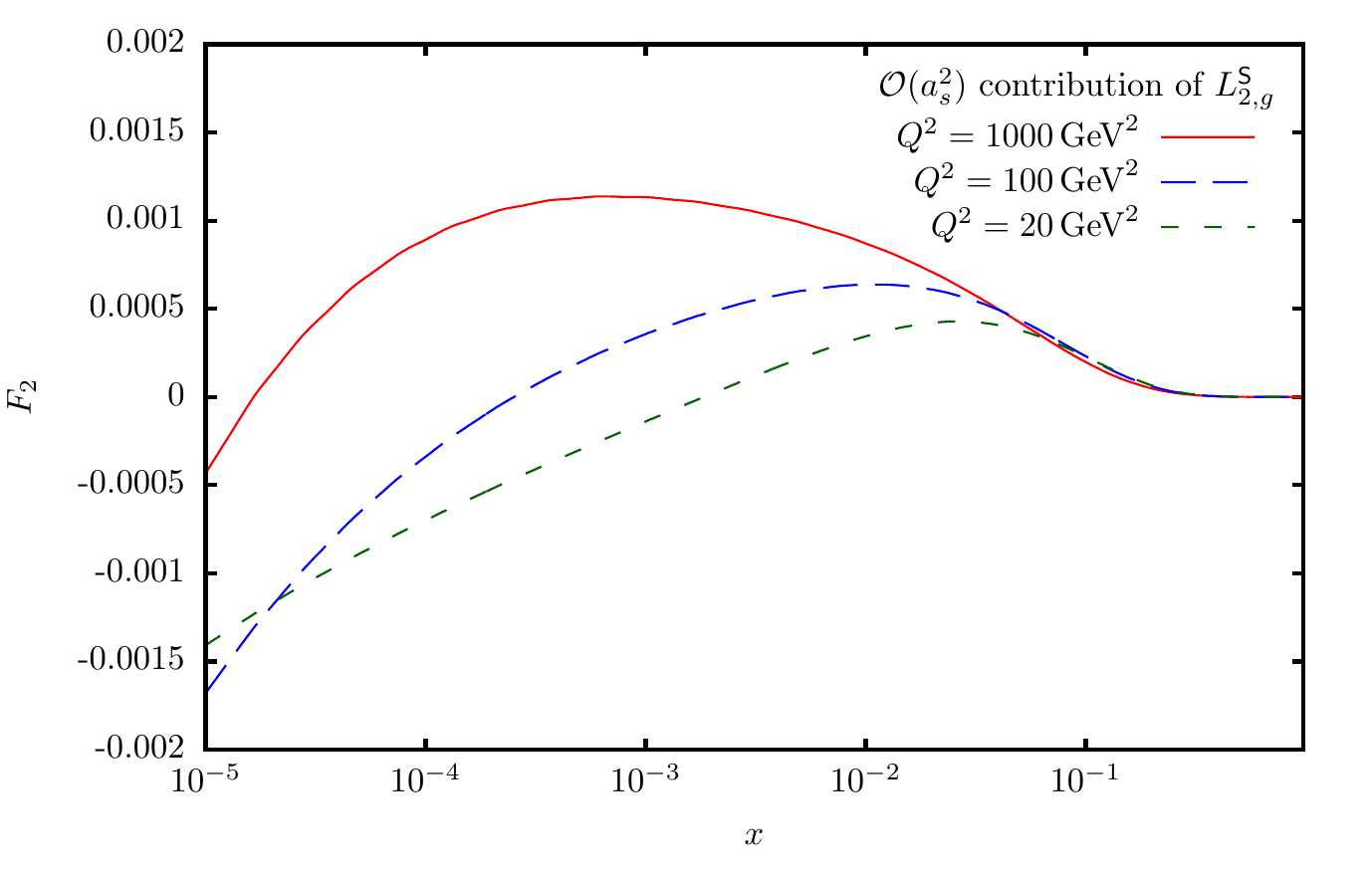}
\includegraphics[width=0.48\textwidth]{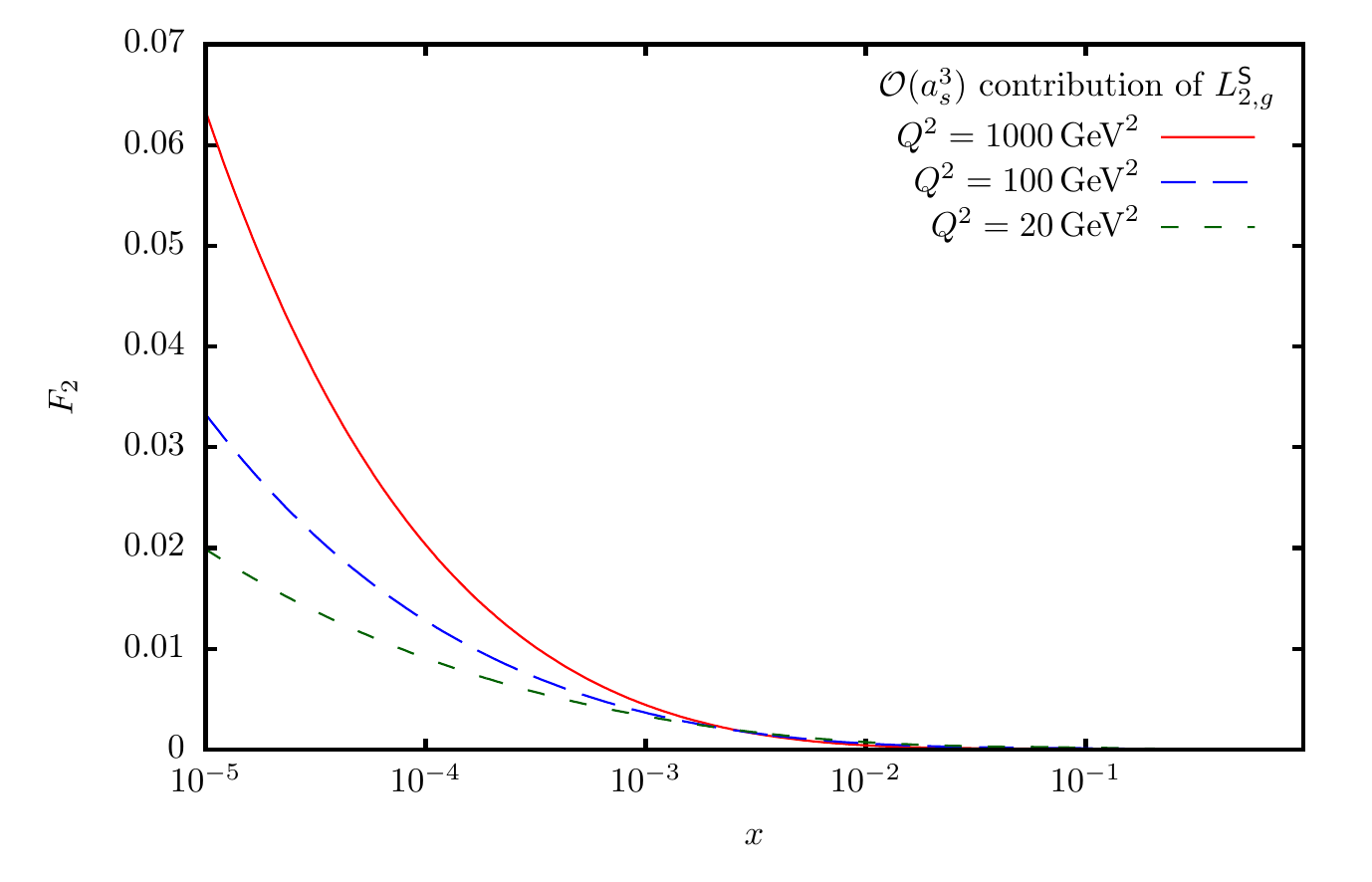}
{\footnotesize \caption[]{
Left panel: the contribution of the Wilson coefficient $L_{2,g}^S$ at $O(a_s^2)$.
Right panel: the $O(a_s^3)$ contributions; from Ref.~\cite{Behring:2014eya}.
}}
\end{figure}

\begin{figure}[h]
\includegraphics[width=0.48\textwidth]{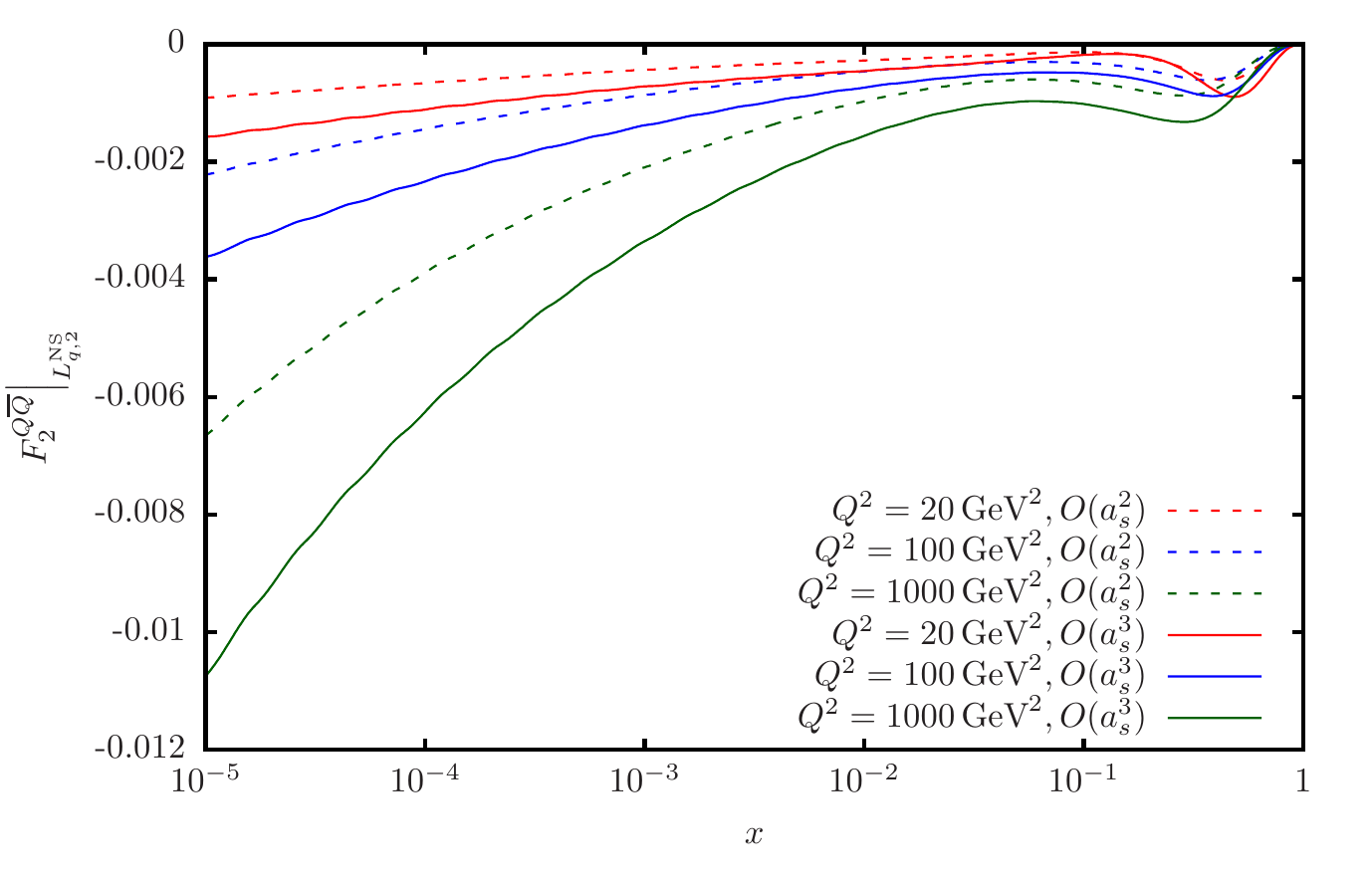}
\includegraphics[width=0.48\textwidth]{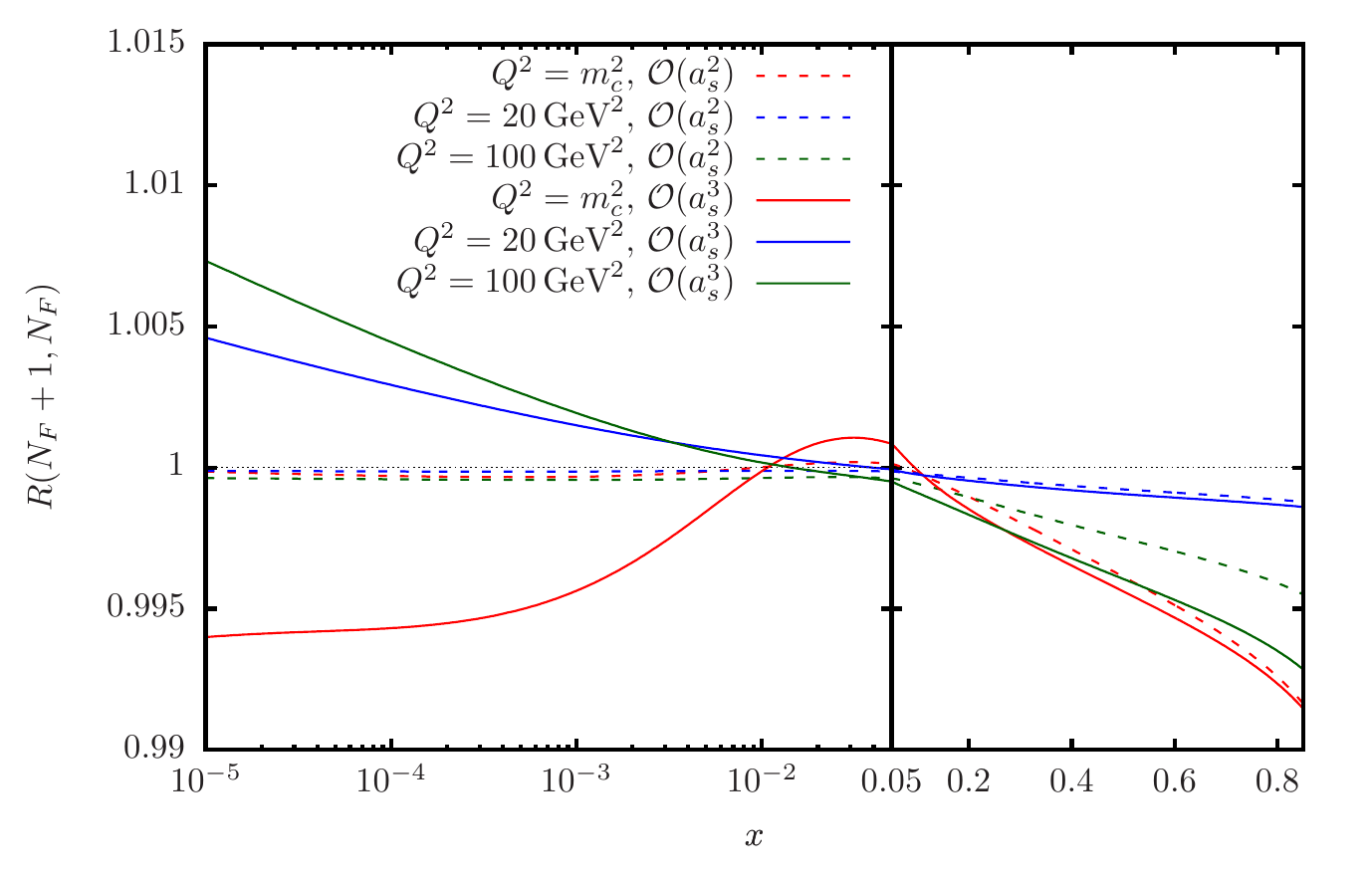}
\caption[]{
Left panel: The massive flavor non-singlet contributions to $F_2(x,Q^2)$.
Right panel: The ratio of the non-singlet $u + \bar{u}$-quark contribution of 4-
to 3 flavors as a function of $x$ for different matching scales; from Ref.~\cite{Ablinger:2014vwa}.
}
\end{figure}
In Figure~2 we illustrate the 2- and 3-loop flavor non-singlet heavy flavor contributions to the structure 
function $F_2(x,Q^2)$ as a function of $x$ and $Q^2$. The 3-loop corrections lower the contributions obtained at 
2-loops. The corrections are of the 1\% level. In the flavor non-singlet case all transition functions in the 
VFNS are available, and we illustrate also the ratio of the non-singlet $u + \bar{u}$-quark contribution of 4- 
to 3 flavors as a function of $x$ for different matching scales. While at small $x$ the 2-loop effects are tiny,
the 3-loop effects show some dependence both at small and larger values of $x$.

Turning to the polarized case, we illustrate in Figure~3 the ratio of the 
charm contribution to the massless contributions to $g_1(x,Q^2)$ as a function of $x$ and $Q^2$ up to 3-loop order. It turns out to 
be positive in the region of lower $x$. The correction becomes negative at large values of $x$ and amounts to a 
few percent. Similarly, we have calculated the 3-loop heavy flavor corrections to the non-singlet structure function
$x(F_3^{W^+} + F_3^{W^-})$. Here the heavy flavor correction varies from $+3\%$ to $-4\%$ from small to large 
$x$.    
\begin{figure}[t]
\includegraphics[width=0.48\textwidth]{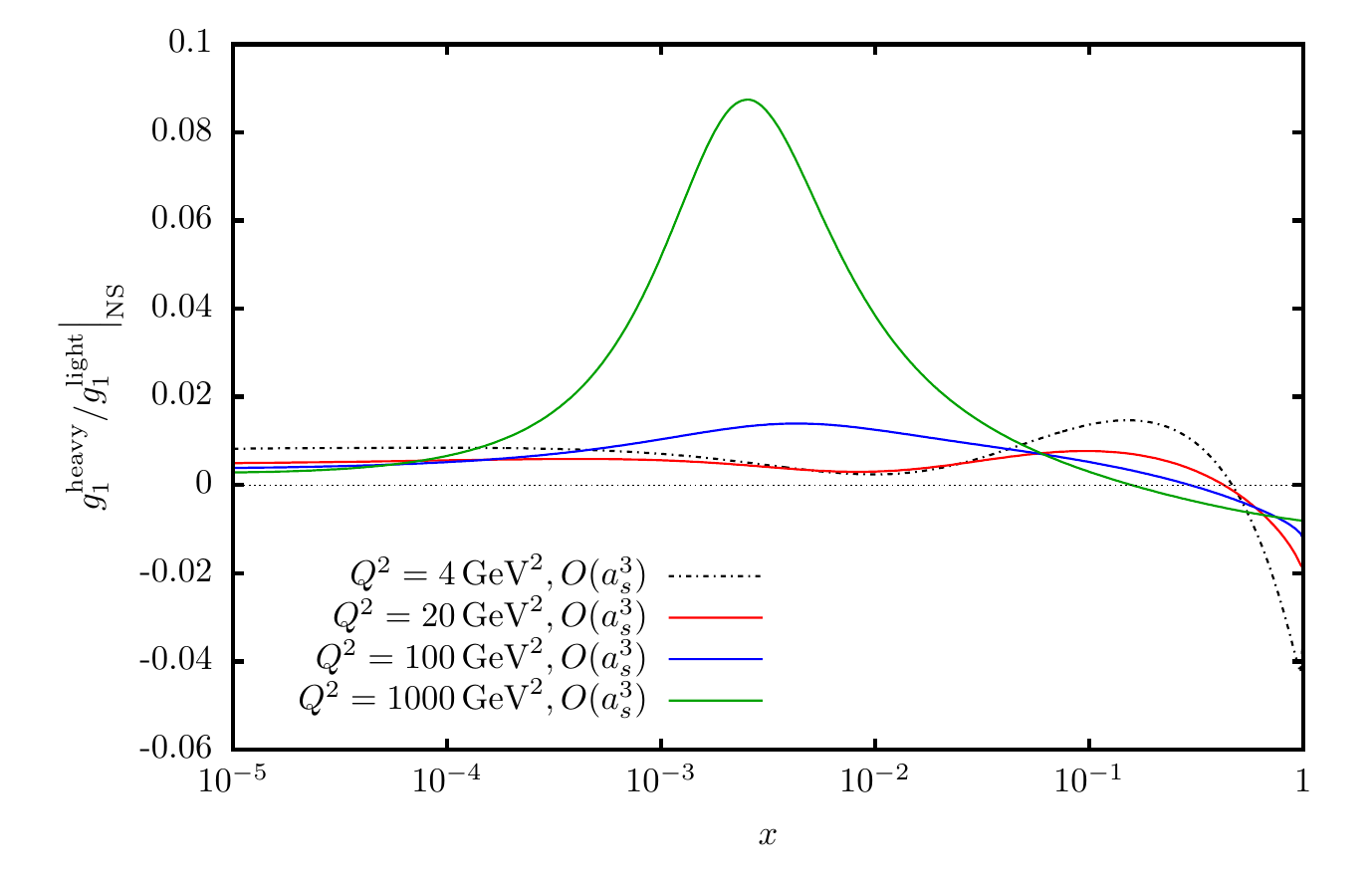}
\includegraphics[width=0.48\textwidth]{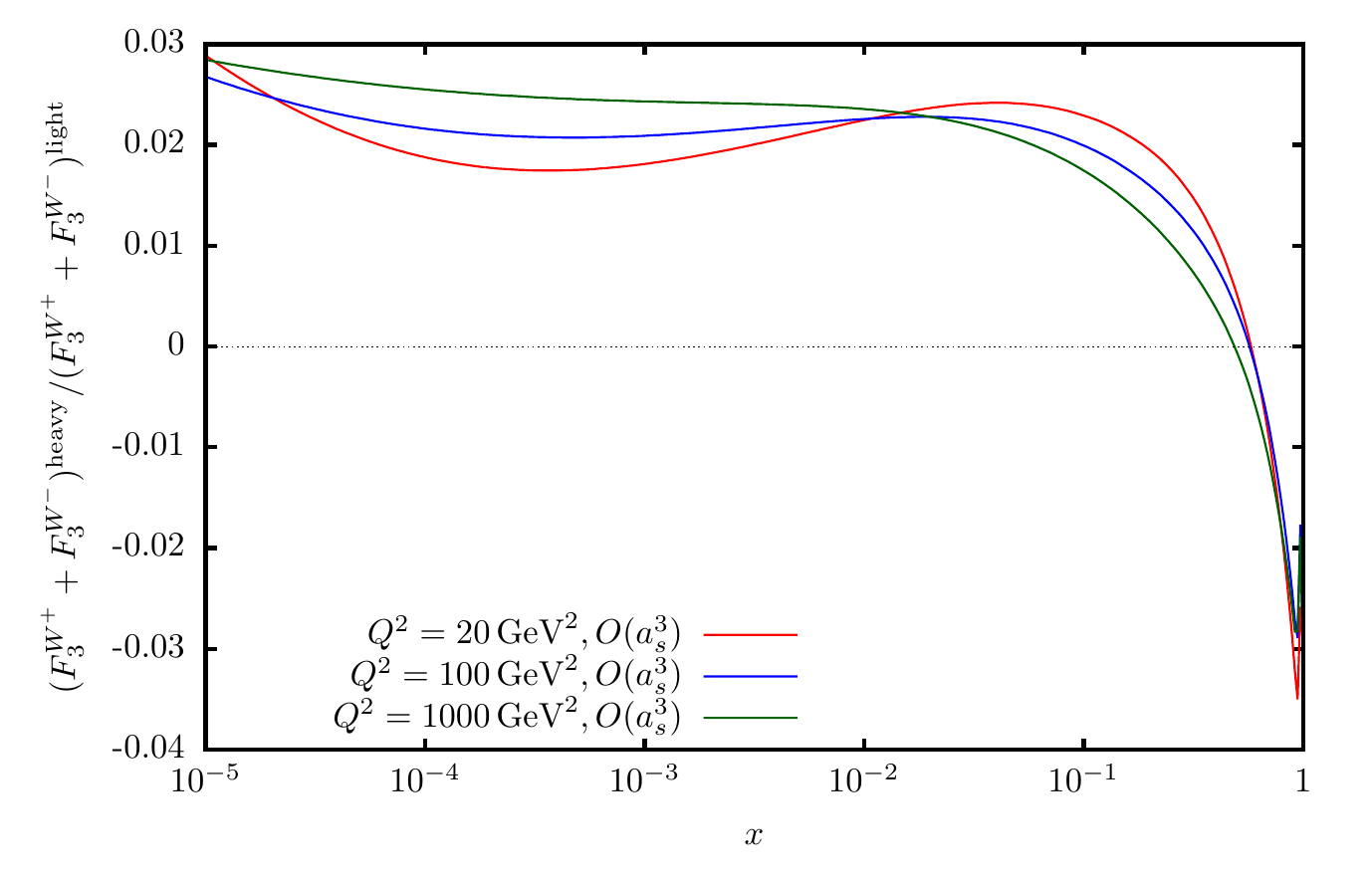}
\caption[]{
Left panel: The ratio of the charm to the massless flavor contribution to the polarized structure function 
$g_1(x,Q^2)$ up to 3-loop order; from Ref.~\cite{Behring:2015zaa}.
Right panel: The ratio of the charm to the massless flavor contribution to the structure function 
$x(F_3^{W^+}+F_3^{W^-})$ up to 3-loop order; from Ref.~\cite{Behring:2015roa}.
}
\end{figure}
\begin{figure}[h]
\includegraphics[width=0.48\textwidth]{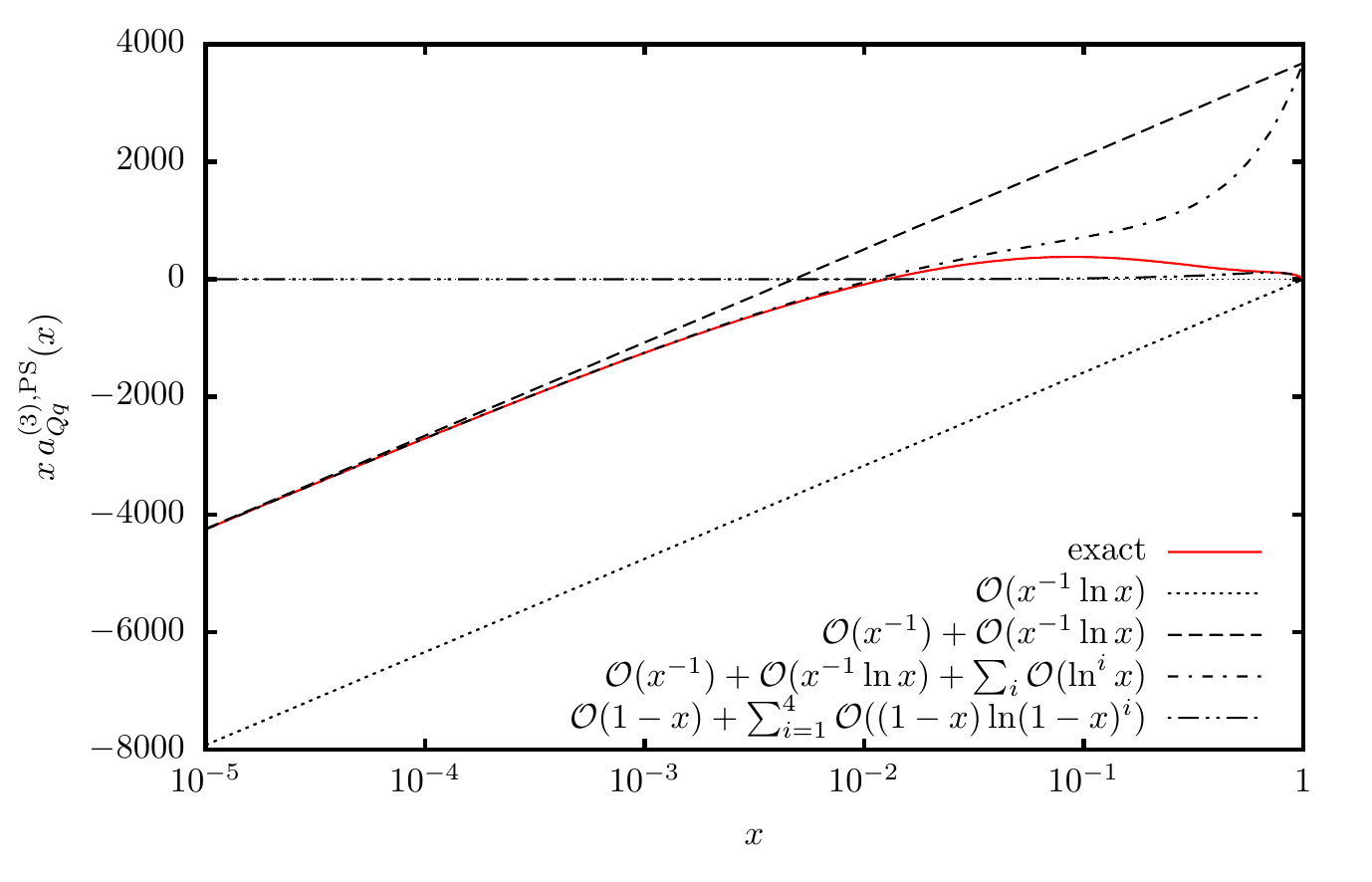}
\includegraphics[width=0.40\textwidth]{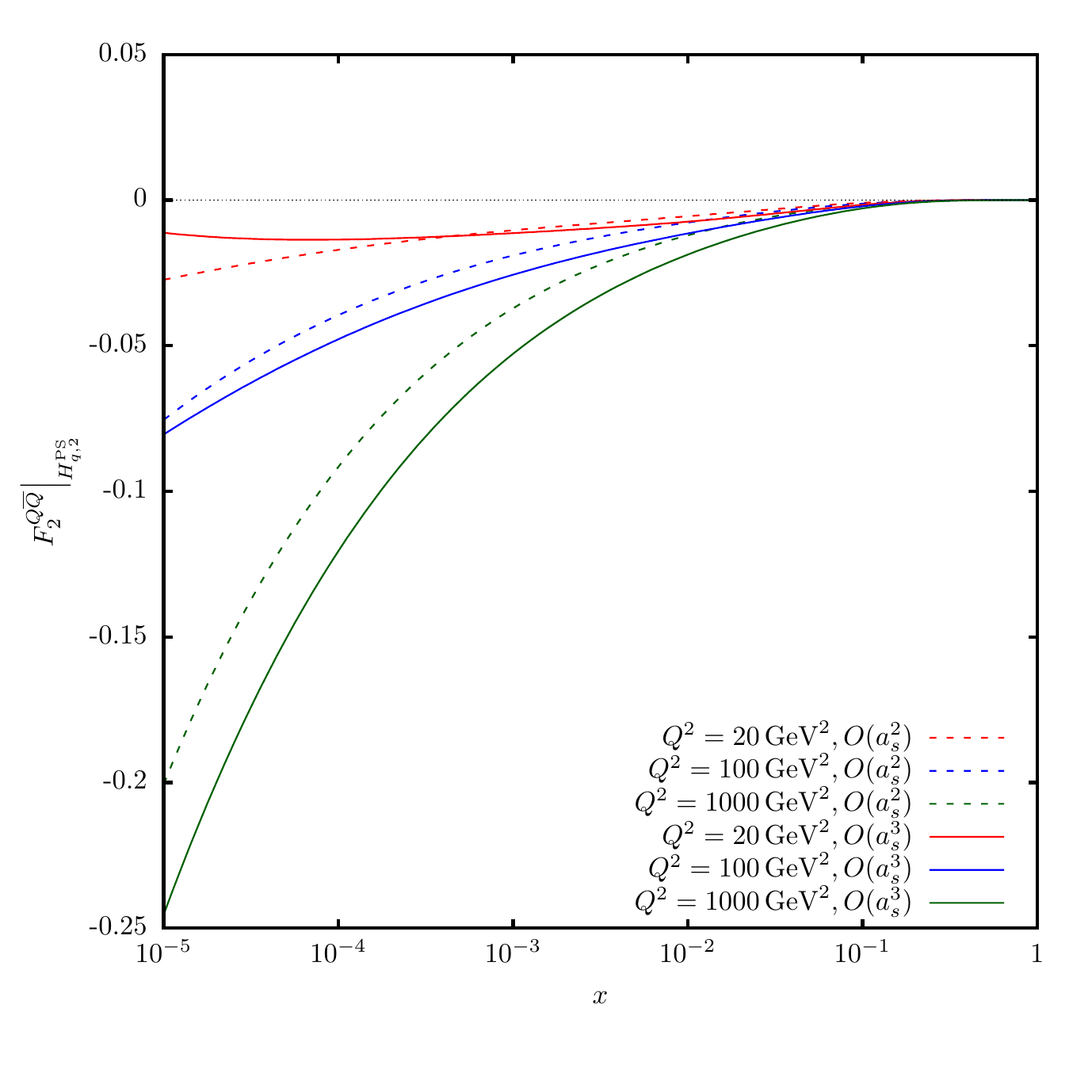}
\caption[]{
Left panel:  $xa_{Qq}^{(3),\rm PS}(x)$ in the low $x$ region (solid red line) and leading terms approximating this
quantity; dotted line: `leading' small $x$ approximation $O(\ln(x)/x)$, dashed line: adding the $O(1/x)$-term, 
dash-dotted line: adding all other logarithmic contributions.
Right panel: Charm pure singlet contribution to $F_2(x,Q^2)$ to $O(a_s^2)$ and $O(a_s^3)$ as a 
function of $Q^2$; from Ref.~\cite{Ablinger:2014nga}.
}
\end{figure}

In Figure~4 we illustrate the 3-loop pure-singlet corrections to the structure function $F_2(x,Q^2)$. First we 
consider the constant part of the unrenormalized OME as a function of $x$. There is a prediction of the 
most singular small $x$ contribution \cite{Catani:1990eg,Kawamura:2012cr}, which is analytically confirmed in 
our calculation. However, this term is misleading and describes this correction nowhere, since subleading terms
are of the same size and are therefore important, cf. \cite{Blumlein:1997em}. The massive 
pure-singlet corrections are larger than those in the non-singlet case and again the 3-loop corrections are 
relevant.
\section{$\mathbf{A_{gg}(N)}$}

\vspace*{1mm}
\noindent
Most recently we have calculated the OME $A_{gg}(N)$ at 3-loop order contributing in the VFNS. In a 
first step we calculated this OME for all its even integer moments for $N \geq 2$. Here the part not 
induced by renormalization and factorization is the constant part of the 3-loop unrenormalized matrix 
element. Its principal structure is given by
\begin{eqnarray}
a_{gg,Q}^{(3)} &=&
\frac{1+(-1)^N}{2}\Biggl\{
{C_F^2 T_F}
\Biggl[\frac{16 \big(N^2+N+2\big)}{N^2 (N+1)^2}
{\sum_{i=1}^N \frac{\binom{2 i}{i} \left(
                                \sum\limits_{j=1}^{i} \frac{4^{j} S_1({j-1})}{\binom{2 j}{j} j^2}
                                -7 \zeta_3
                        \right)}{4^i \big(i+1\big)^2}}
 \nonumber\\ &&
                -\frac{4 P_{69} S_1^2}{3 (N-1) N^4 (N+1)^4 (N+2)}
                +\tilde{\gamma}^{(0)}_{gq} \Biggl(
                        \frac{128 \left(S_{-4} - S_{-3} S_1 + S_{-3,1} + 2 S_{-2,2}\right)}{3 N
(N+1) (N+2)}
\nonumber \\ &&
                        +\frac{4 \big(5 N^2+5 N-22\big) S_1^2 S_2}{3 N (N+1) (N+2)} + \cdots
\Biggr) + \cdots \Biggl]
\nonumber \\ &&
        +{C_A C_F T_F} \Biggl[
                \frac{16 P_{42}}{3 (N-1) N^2 (N+1)^2 (N+2)}
                        {\sum_{i=1}^N \frac{\binom{2 i}{i} \left(
                                \sum\limits_{j=1}^{i}\frac{4^{j} S_1({j-1})}{\binom{2 j}{j} j^2}
                                -7 \zeta_3
                        \right)}{4^i \big(i+1\big)^2}}
\nonumber \\ &&
                +\frac{32 P_2 S_{-2,2}}{(N-1) N^2 (N+1)^2 (N+2)}
                -\frac{64 P_{14} S_{-2,1,1}}{3 (N-1) N^2 (N+1)^2 (N+2)}
\nonumber \\ &&
                -\frac{16 P_{23} S_{-4}}{3 (N-1) N^2 (N+1)^2 (N+2)}
                +\frac{4 P_{63} S_4}{3 (N-2) (N-1) N^2 (N+1)^2 (N+2)} + \cdots \Biggr]
\nonumber \\ &&
        +{C_A^2 T_F} \Biggl[
                -\frac{4 P_{46}}{3 (N-1) N^2 (N+1)^2 (N+2)}
                        {\sum_{i=1}^N \frac{\binom{2 i}{i} \left(
                                \sum\limits_{j=1}^{i} \frac{4^{j} S_1({j-1})}{\binom{2 j}{j} j^2}
                                -7 \zeta_3
                        \right)}{4^i \big(i+1\big)^2}}
\nonumber \\ &&
                +\frac{256 P_5 S_{-2,2}}{9 (N-1) N^2 (N+1)^2 (N+2)}
                +\frac{32 P_{30} S_{-2,1,1} + 16 P_{35} S_{-3,1} + 16 P_{44} S_{-4}}{9 (N-1) N^2
(N+1)^2 (N+2)}
\nonumber \\ &&
                 +\frac{16 P_{52} S_{-2}^2}{27 (N-1) N^2 (N+1)^2 (N+2)}
                +\frac{8 P_{36} S_2^2}{9 (N-1) N^2 (N+1)^2} + \cdots \Biggr]
\nonumber \\ &&
        +{C_F T_F^2} \Biggl[
                -\frac{16 P_{48} {\binom{2 N}{N} 4^{-N} \left(
                        \sum_{i=1}^N \frac{4^{i} S_1({i-1})}{\binom{2 i}{i} i^2}
                        - 7\zeta_3
                \right)}}{3 (N-1) N (N+1)^2 (N+2) (2 N-3) (2 N-1)}
\nonumber \\ &&
                -\frac{32 P_{86} S_1}{81 (N-1) N^4 (N+1)^4 (N+2) (2 N-3) (2 N-1)}
\nonumber \\ &&
                +\frac{16 P_{45} (S_1^2 - S_2/3)}{27 (N-1) N^3 (N+1)^3 (N+2)}
+ \cdots \Biggr] + \cdots
\Biggr\},
\end{eqnarray}
displaying some of the new nested sums, with $P_j$ polynomials in $N$. The contributions $\propto N_F 
T_F^2$ or $T_F^2$ were calculated before. Also, with this calculation we have rederived all the 
contributions to the 3-loop anomalous dimensions $\propto T_F$.
\section{Analytic Two-Mass Contributions}

\vspace*{1mm}
\noindent
In the previous sections we have discussed the case of $N_F$ massless and one massive quark. Starting with 3-loop order, 
there are also graphs that contain two massive lines of different mass. The usual VFNS, 
cf. e.g. \cite{Bierenbaum:2009mv},
is therefore no longer applicable and one cannot define individual heavy flavor parton densities anymore. This is 
due to the fact that the ratio $\eta = m_c^2/m_b^2$ is $\sim 1/10$ only and charm cannot be considered to be massless already at 
the bottom 
mass scale. In \cite{TWOM1,FWTHESIS} we have calculated the moments $N = 2,4,6$ for all contributing OMEs to a 
precision of 
better than $10^{-4}$ using the package {\tt q2e} \cite{Q2E}. The complete results in the non-singlet case and 
for all contributing scalar integrals to $A_{gg}^{(3)}$ 
have been calculated in \cite{TWOM2,FWTHESIS}. Here various new iterated integrals over root-valued letters 
appear with
also the variable $\eta$ as parameter. Calculating these corrections for general values of the Mellin variable $N$, it is not 
always possible to expand in the parameter $\eta$. We therefore have calculated the complete corrections analytically in these 
cases. 
\section{Conclusions}

\vspace*{1mm}
\noindent
In 2009
a series of Mellin moments up to $N = 10-14$ for all massive 3--loop OMEs and Wilson coefficients have been 
calculated, followed by the computation of the first two massive 3-loop Wilson coefficients for $F_2(x,Q^2)$, 
$L_q^{\rm (3), PS}(N)$ and $L_g^{\rm (3), S}(N)$, in the asymptotic region in 2010, completing soon after all 
contributions $\propto N_F T_F^2$. The logarithmic contributions are completely known. Ladder, $V$-Graph and 
Benz-topologies for graphs, with no singularities in $\varepsilon$, were systematically calculated for general 
values of $N$ in 2013. The corresponding case of singular structures
and associated physical diagrams has been solved in 2015. Here new functions occur, including iterated integrals  
with a larger 
number of root-valued letters of quadratic forms. In 2014 we calculated, based on the technologies described in 
Section~\ref{sec:3},
the 3-loop Wilson coefficients and massive OMEs $L_q^{\rm NS, (3)}$, $A_{gq,Q}^{\rm S, (3)}$, $A_{qq,Q}^{\rm NS, TR (3)}$,
{$H_{2,q}^{\rm PS (3)}$} and {$A_{Qq}^{\rm PS (3)}$}. A method for the calculation of graphs with two massive lines
of equal masses and operator insertions has been developed and applied to $A_{gg,Q}^{(3)}(N)$. The method can 
be generalized to the case of unequal masses. 
Here the moments for $N=2,4,6$ for all graphs with two quark lines of unequal masses are now known, which requires extensions in the
renormalization procedure. It turns out in general, that in the case of general values of $N$ one has to compute the complete result,
as the expansion in $m_c^2/m_b^2$ not always commutes with the functional structure in $N$, unlike the case of fixing $N$ a priori. 
We also calculated the charged current Wilson coefficients up to $O(\alpha_s^2)$ correcting some errors in the foregoing literature. 

In all cases the corresponding 3--loop anomalous dimensions were computed as functions of $N$ and $x$ resp., those for
{transversity} for the first time ab initio. The corresponding contributions are $\propto N_F^k, k\geq 1$, the latest example being
$\hat{\gamma}_{gg}^{(3)}$. In the pure singlet case this is the first complete recalculation using a different method than in 
Ref.~\cite{Vogt:2004mw}.
All master integrals for {$A_{gg,Q}^{(3)}$} have been computed and the complete expression for general values of $N = 2k, k\in 
\mathbb{N}, k \geq 1$ has been calculated in 2015. The computation of the OME {$A_{Qg}^{(3)}(N)$} made also 
some progress. Here all
master integrals that can be thoroughly represented as nested sums in difference fields have been calculated. 
A series of new computer-algebra and mathematical technologies were developed and encoded in the still growing packages 
{\tt Sigma, EvaluateMultiSums, SumProduction, SolveCoupledSystem, $\rho$Sum, HarmonicSums} and {\tt MultiIntegrate} 
for efficient use.
In the present project still more involved structures await their analytic solution. As for those structures having 
been found in
the past we expect that they also appear in other massive higher-loop precision calculations in collider physics. 

\vspace{5mm}
\noindent
{\bf Acknowledgment.}~
We would like to thank M.~Steinhauser for the possibility to use the package {\tt MATAD3.0}. 


\end{document}